\documentclass{emulateapj}

\usepackage{scalerel,tikz}
\usetikzlibrary{svg.path}
\definecolor{orcidlogocol}{HTML}{A6CE39}
\tikzset{orcidlogo/.pic={
 \fill[orcidlogocol] svg{M256,128c0,70.7-57.3,128-128,128C57.3,256,0,198.7,0,128C0,57.3,57.3,0,128,0C198.7,0,256,57.3,256,128z};
 \fill[white] svg{M86.3,186.2H70.9V79.1h15.4v48.4V186.2z}
 svg{M108.9,79.1h41.6c39.6,0,57,28.3,57,53.6c0,27.5-21.5,53.6-56.8,53.6h-41.8V79.1z M124.3,172.4h24.5c34.9,0,42.9-26.5,42.9-39.7c0-21.5-13.7-39.7-43.7-39.7h-23.7V172.4z}
 svg{M88.7,56.8c0,5.5-4.5,10.1-10.1,10.1c-5.6,0-10.1-4.6-10.1-10.1c0-5.6,4.5-10.1,10.1-10.1C84.2,46.7,88.7,51.3,88.7,56.8z};
}}
\newcommand\orcidicon[1]{\href{https://orcid.org/#1}{\mbox{\scalerel*{
\begin{tikzpicture}[yscale=-1,transform shape]
\pic{orcidlogo};
\end{tikzpicture}
}{|}}}}

\def\ga{\mathrel{\hbox{\rlap{\hbox{\lower4pt\hbox{$\sim$}}}\hbox{$>$}}}}
\def\la{\mathrel{\hbox{\rlap{\hbox{\lower4pt\hbox{$\sim$}}}\hbox{$<$}}}}

\usepackage{verbatim}
\usepackage{amsmath}
\usepackage{multirow}
\usepackage{relsize}
\usepackage{color}
\usepackage[colorlinks,citecolor=blue]{hyperref}

\usepackage{amssymb}
\usepackage{float}

\newcommand{\sigs}{\sigma_s}

\newcommand{\deriv}{\,\mathrm{d}}

\newcommand{\ls}{\lambda_{\rm s}}

\newcommand{\lj}{\lambda_{\rm J}}

\newcommand{\cs}{c_{\rm s}}

\shorttitle{The Dense Gas Fraction and Star Formation}

\shortauthors{Burkhart \& Mocz}
\begin{document}
\title{The Self-gravitating Gas Fraction and The Critical Density for Star Formation}
\author{Blakesley Burkhart~\orcidicon{0000-0001-5817-5944}\altaffilmark{1,2}\& Philip Mocz~\orcidicon{0000-0001-6631-2566} \altaffilmark{3,$\dagger$}}
\altaffiltext{1}{Center for Computational Astrophysics, Flatiron Institute, 162 Fifth Avenue, New York, NY 10010, USA}
\altaffiltext{2}{Department of Physics and Astronomy, Rutgers University,  136 Frelinghuysen Rd, Piscataway, NJ 08854, USA}
\altaffiltext{3}{Department of Astrophysical Sciences, Princeton University, 4 Ivy Lane, Princeton, NJ, 08544, USA}
\altaffiltext{$\dagger$}{Einstein Fellow}

\begin{abstract}
 We analytically calculate the star formation efficiency and dense self-gravitating gas fraction in the presence of  magneto-gravo-turbulence using the model of Burkhart (2018), which employs a piecewise lognormal \textit{and} powerlaw density Probability Distribution Function (PDF). We show that the PDF transition density from lognormal to powerlaw forms is a mathematically motivated critical density for star formation and can be physically related to the density where the Jeans length is comparable to the sonic length, i.e. the post-shock critical density for collapse. When the PDF transition density is taken as the critical density, the instantaneous star formation efficiency ($\epsilon_{\rm inst}$) and depletion time ($\tau_{\rm depl}$) can be calculated from the dense self-gravitating gas fraction represented as the fraction of gas in the PDF powerlaw tail. 
 We minimize the number of free parameters in the analytic expressions for $\epsilon_{\rm inst}$ and $\tau_{\rm depl}$ by using the PDF transition density instead of a parameterized critical density for collapse and thus provide a more direct pathway for comparison with observations. We test the analytic predictions for the transition density and self-gravitating gas fraction against AREPO moving mesh gravoturbulent simulations and find good agreement.  We predict that, when gravity dominates the density distribution in the star forming gas, the star formation efficiency should be weakly anti-correlated with the sonic Mach number while the depletion time should increase with increasing sonic Mach number. The star formation efficiency and depletion time depend primarily on the dense self-gravitating gas fraction, which in turn depends on the interplay of gravity, turbulence and stellar feedback. Our model prediction is in agreement with recent observations, such as the M51 PdBI Arcsecond Whirlpool Survey (PAWS).

\end{abstract}

\section{Introduction}
\label{intro}

Star formation in galaxies depends on the complex relationship between gravity, magnetic fields, feedback and partially ionized fluid motions.  Turbulence and magnetic fields contribute to the overall inefficiency of galaxies to convert gas into stars, however an exact theoretical description of how these processes connect to the observed star formation rates in galaxies and the initial mass function (IMF) remains mysterious \citep{krumreview2014}.  Nevertheless, observational studies indicate that the most important factor for predicting star formation rates in galaxies is the amount of dense gas which is unstable to gravitational collapse \citep{Lada10a,Lada2012}.

How does diffuse molecular and atomic gas in galaxies become dense and collapse to form stars?
Most analytic star formation theories rely on supersonic turbulence to produce gravitationally unstable density fluctuations as well as set the overall fraction of dense star formation gas. The density distribution expected for supersonic magnetized isothermal turbulence  is a lognormal \citep{Vazquez-Semadeni1994,Vazquez-Semadeni95a,Padoan1997,Scalo98a,Kravtsov03a,Robertson2008,Hennebelle08b,Price10b,Collins12a,Burkhart12,Hopkins13b,Walch13a}.

Deviations from lognormal are expected for different equations of state \citep{federrath2015MNRAS.448.3297F}, but for isothermal turbulence the width of the lognormal ($\sigs$) is given by the turbulence sonic Mach number $M_s =v_L/\cs$ \citep{Krumholz2005,Federrath2008,Burkhart09}, which depends on the rms velocity dispersion ($v_L$), the sound speed ($\cs$) and turbulence driving parameter $b$:
\begin{equation}
\sigs^2=\ln[1+b^2M_s^2]
\label{eqn.sigma}
\end{equation}

Most previous analytic calculations for the star formation rate have all hinged on the form of the density PDF being fully lognormal \citep{Krumholz2005,Padoan11b,Hennebelle11b,federrath12,Renaud12a,Hopkins12d,Gribel2017}.  These works calculate the star formation rate (SFR) by integrating the lognormal PDF from a critical density for collapse, which varies for different authors. In these works, the SFR depends on the exact choice of a number of parameters of order unity and the width of the lognormal PDF (e.g. given by Equation~\ref{eqn.sigma}). In most cases, the expected dependence of the star formation rate per free fall time (SFR$_{\rm ff}$) on the sonic Mach number is that the SFR$_{\rm ff}$ should increase
with increasing $M_s$ or increasing compression factor (b) because higher Mach number implies
stronger local compression \citep{Padoan11b,Hennebelle11b,federrath12}.  The exception to this is the \citet{Krumholz2005} formalism for low viral parameters ($\alpha_{\rm vir}$) where SFR$_{\rm ff}$ stays constant with larger sonic Mach number (See Figure 1 of \citet{federrath12}). Recent observations have shown that the star formation efficiency per free fall time $\epsilon_{\rm ff}$ may be anti-correlated with sonic Mach number, which is in tension with some of the above theories \citep{Leroy2017}.

Observational and numerical work has determined that the dense gas in molecular clouds (i.e. extinctions greater than $A_v >1$ and/or $n\approx10^3$cm$^{-3}$) is predominately found to have a powerlaw PDF rather than a lognormal form \citep{Kritsuk11b,Collins12a,Girichidis2014,MyersP2015,schneider2015MNRAS.453L..41S,Lombardi2015AA,Burkhart2017ApJ...834L...1B,Mocz2017,padoan2017ApJ...840...48P,MyersP2017,Chen2017,Alves2017AA}.  If the density PDF of molecular gas is not lognormal, this may provide an avenue with which to relieve observational tension with the above mentioned analytic models \citep{Burkhart2018}.
Observational studies in particular have been instrumental in dissecting the density distribution in and around giant molecular clouds (GMCs) and have confirmed that the highest column
density regime (corresponding to visual extinction A$_V >1$) of the PDF often has a powerlaw distribution while the lower column density material in the PDF (traced by diffuse molecular and atomic gas) is well-described by a lognormal form \citep{Wada07a,Kainulainen09a,Lombardi10a,schneider2015MNRAS.453L..41S,schneider15,Kainulainen13a,Hennebelle11a,federrath12,Kainulainen2014,Stutz2015A&A...577L...6S,Lombardi2015AA,Burkhart2015,Imara2016,Bialy2017ApJ...843...92B,Chen2017,Kainulainen2017,Scannapieco2018}.
In this paper we calculate the star formation efficiency based on the formalism first presented in \citet{Burkhart2018}, where the PDF was considered as a piecewise lognormal and powerlaw with the transition happening at an analytically determined transition density, which was derived and tested in a series of observational and numerical works \citep{Collins12a,burkhartcollinslaz2015,Burkhart2017ApJ...834L...1B,Chen2017}.  \citet{Burkhart2018} considered a molecular cloud which was undergoing collapse in the very initial stages of star formation, i.e. within the first cloud mean free fall time. There the star formation rate was found to rapidly accelerate past the predictions for the lognormal only density PDF calculation, again in agreement with numerical and observational works \citep{Murray11b}. 
The powerlaw slope shallows significantly in less than the mean free fall time, while the lifetimes of GMCs are typically between $2$--$10$ free fall times \citep{Palla00a,Meidt2015,jefferson2018,Kruijssen2018}.  Therefore, the powerlaw portion of the PDF can not be ignored in the SFR calculation \citep{Burkhart2018} and is related to the observed accelerated rates of star formation \citep{Murray11b,Lee2015ApJ...800...49L,Lee2016}. \citet{Girichidis2014} and \citet{Guszejnov2018} found analytically that for collapsing clouds the powerlaw slope of the density PDF should saturate to values between $\alpha=1$--$1.5$ which was also confirmed by numerical studies without feedback \citep{Collins12a,Lee2015ApJ...800...49L,burkhartcollinslaz2015,Mocz2017}. However, the picture is complicated by influence of stellar feedback, which may steepen the powerlaw slope and reduce the overall star formation efficiency \citep{Federrath2015,Federrath2016,Semenov2017,Grudic2018}.

 In this paper, we will show how the use of the PDF transition density allows us to largely eliminate unobservable free parameters from the  critical density for star formation. As we will show, this allows for a calculation of the dense self-gravitating gas fraction, star formation efficiency and gas depletion time which can more readily be compared with observations.
This paper is organized as follows:
In Section ~\ref{section.math} we review the piecewise lognormal and powerlaw transition density calculation and derive the normalization shift density for situations which require mass conservation, e.g. as is the case with most turbulent box simulations. 
In Section~\ref{sec:trans} we show that the PDF transition density between lognormal and powerlaw forms is a mathematically motivated critical density and can be physically related to the density where the Jeans length is comparable to the sonic length, i.e. the post-shock critical density for collapse.  We test these results with {\sc AREPO} gravoturbulent simulations with different sonic Mach numbers. 
In Section~\ref{section.SFE} we use the star formation rate model of \citet{Burkhart2018} to calculate the self-gravitating gas fraction for active star forming clouds. We use this analytic form for the self-gravitating gas fraction to compute the star formation efficiency (SFE). In Section~\ref{section.obs} we compare the model SFE to observations from the PdBI Arcsecond Whirlpool Survey (PAWS, \citet{Schinnerer13a}), which include CO derived velocity dispersion information.  This allows us to test a number of predictions, namely that the star formation efficiency is slightly anti-correlated with sonic Mach number for actively star forming GMCs with powerlaw PDFs. In Section~\ref{section.sum} we  discuss our results followed by our conclusions in Section~\ref{sec:con}.

\section{The Form of the Density PDF in a Gravoturbulent Medium}
\label{section.math}
We consider a piecewise form of the density PDF ($p_{LN+PL}(s)$) in and around a star forming molecular cloud that consists of a lognormal at low density, a powerlaw at high density and a transition point  ($s_t={\rm ln}(\rho_t/\rho_0)$) between the two \citep{Collins12a,burkhartcollinslaz2015,Burkhart2017ApJ...834L...1B,Burkhart2018}. Here $\rho_t$ is the transition density and $\rho_0$ is the mean density.  Our analytic setup is identical to \citet{Burkhart2018}, their section 3, and therefore we review a few of the key equations here and refer the reader to Paper 1 for a more detailed derivation.

In particular, the transition point for the density takes a simple form:
\begin{align}
s_t&=(\alpha -1/2)\sigs^2
\label{eqn.st}
\end{align}

where $\alpha$ is the slope of the powerlaw tail and $\sigs$ is the lognormal width (e.g, given by Equation~\ref{eqn.sigma}).

\begin{figure*}
\centering
\includegraphics[width=0.97\textwidth]{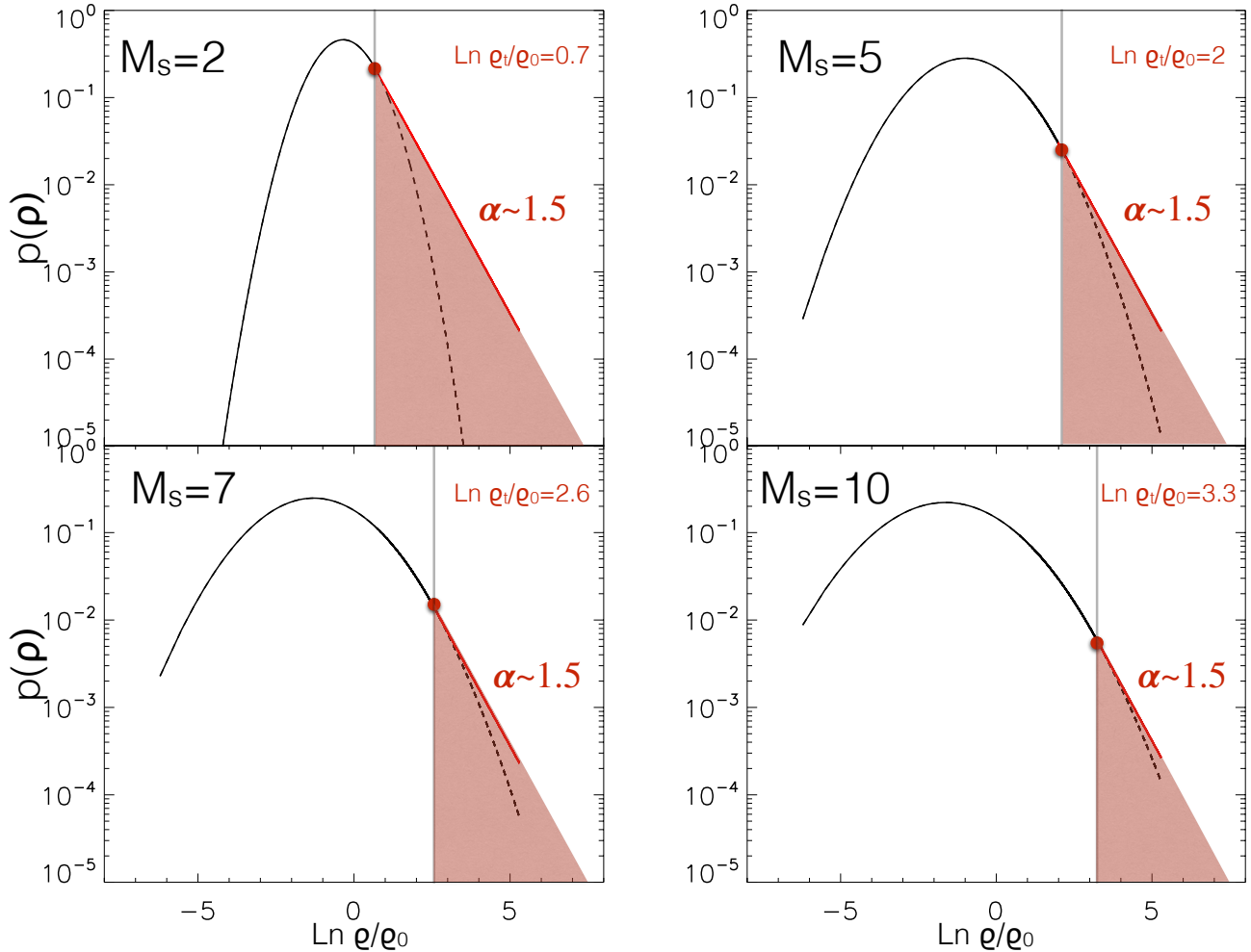}
\caption{
\label{fig:pdf_theory}
Four piecewise lognormal plus powerlaw models with different sonic Mach numbers in each panel.  The Sonic Mach number controls the width of the lognormal, with large sonic Mach number corresponding to larger width of the PDF (Equation~\ref{eqn.sigma}). Each panel has a powerlaw slope of $\alpha=1.5$, which is expected for strongly self-gravitating high density regions. The red box outlines all the density past the transition density (Equation~\ref{eqn.st}, denoted by red dot and black vertical line) which is the dense self-gravitating gas. Gas {below} the transition density in the lognormal is diffuse and potentially unbound \citep{Chen2018}. The value of the transition density  (denoted in each panel) is self-consistently determined by the properties of the PDF, namely the slope of the powerlaw and the width of the lognormal.}
\end{figure*}

As $\alpha$ becomes less steep, the transition density ($s_t$) between the PDF lognormal and powerlaw moves towards lower density (Equation~\ref{eqn.st}), assuming the properties of the turbulence remain the same and therefore $\sigs$ is roughly constant.

\begin{figure}
\centering
\includegraphics[width=0.5\textwidth]{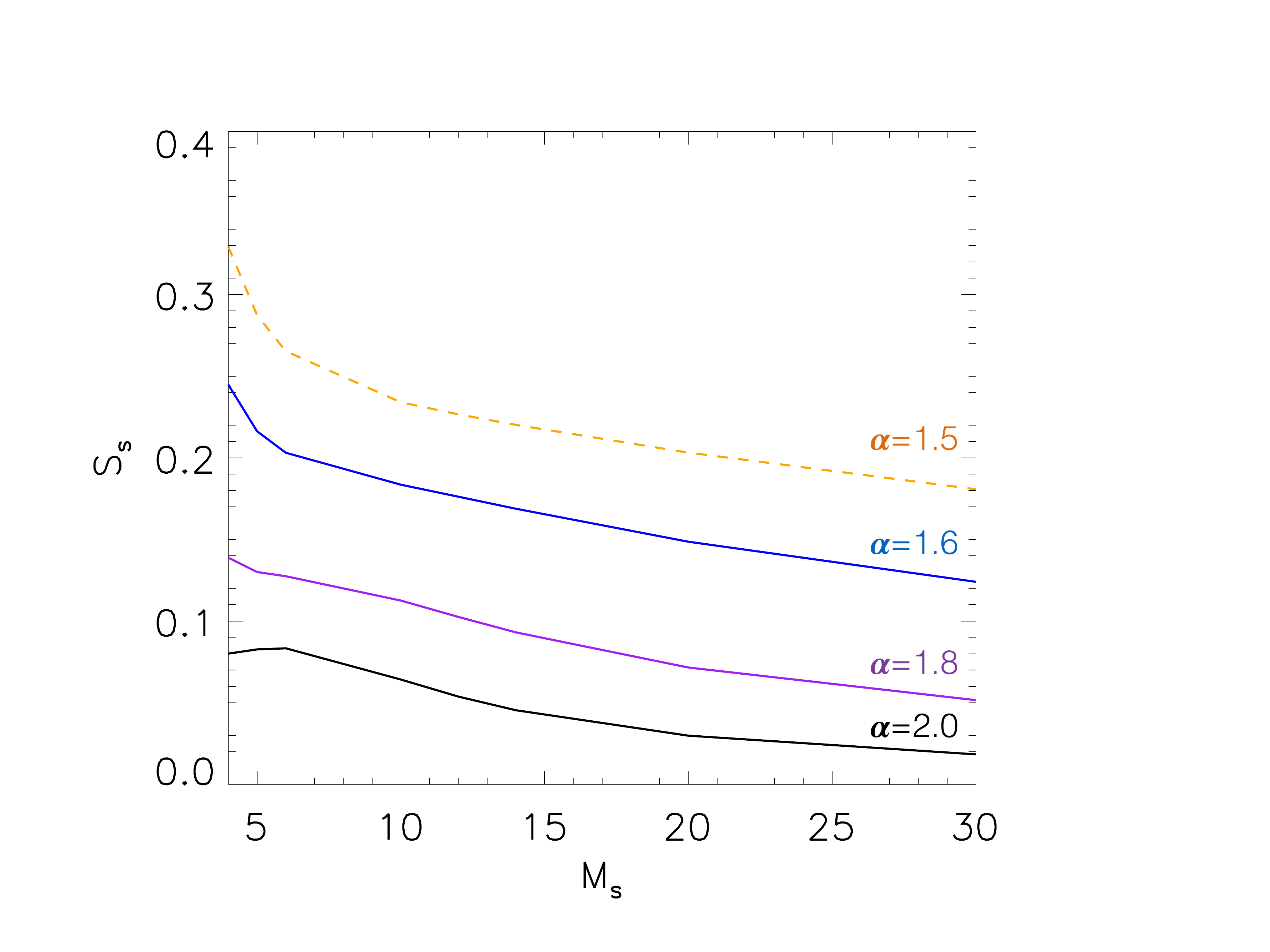}
\caption{
\label{fig:shift}
The density shift vs. sonic Mach number as given by Equations~\ref{eqn.sigma} and~\ref{eqn.shift}. Different values of the powerlaw slope are shown in different colored lines with labels on the far right side of the plot.  The PDF of a gravoturbulent medium saturates to $\alpha \approx 1.5$, which we indicate with the dashed line. At later stages in the cloud evolution, feedback may steepen the PDF slope again.}
\end{figure}

Numerical simulations of gravoturbulence suggest that the PDF of non-collapsing regions retrain the characteristics of 
the initial supersonic turbulence field (e.g. remain lognormal) while 
the density PDFs of collapsing
regions show a clear powerlaw at high density \citep{Kritsuk11b,Collins12a,Lee2015,Mocz2017}. Once the critical density for gravitational collapse is reached the powerlaw begins to form.  The characteristic slope of that powerlaw changes in roughly the cloud mean free fall time from steep, $\alpha(t) \approx 3$, to shallow values, $\alpha(t)\approx1.5$--$1$ \citep{Girichidis2014,Burkhart2017ApJ...834L...1B,Guszejnov2018}. The overall value of $\alpha(t)$ depends on the strength of the magnetic field \citep{burkhartcollinslaz2015} and the efficiency of feedback \citep{2015MNRAS.450.4035F,padoan2017ApJ...840...48P}. 
We also point out that gas with a spherical density profile of $\rho\propto r^{-\kappa}$ has a corresponding PDF for $s$ which is a powerlaw that scales as ${\rm e}^{-\alpha s}$ with $\kappa=3/\alpha$, e.g. \citep{Kainulainen2017}. Hence, a powerlaw distribution of gas with $\alpha=1.5$ is consistent with all the high density gas having collapsed into isothermal cores: $\rho\propto r^{-2}$ \citep{shu77}.

An additional important point to stress is that as the sonic Mach number and/or the compressibility of the medium increases, $s_t$ moves towards higher density since the PDF widens. This implies there is more unbound material overall when the powerlaw tail is included for a cloud with given mean density
\footnote{Despite the different form of the PDF used, this is also the case in the model of \citet{Krumholz2005}. Increasing $M_s$
increases the critical density for collapse and raises the over-density that the gas must
reach to collapse. At the same time, however, increasing
the sonic Mach number increases the width of the probability distribution
function, putting a larger fraction of the gas at high over-density. These two effects produce a slight anti-correlation with Mach number, similar to the model presented in Section 4, despite the different PDF forms used.}. We illustrate the change of the powerlaw slope and transition density in Figure~\ref{fig:pdf_theory}. Comparing the top left panel with $M_s=2$ to the bottom right panel with $M_s=10$ shows this effect. 

Finally, we note that the PDF has been normalized to a pre-collapse reference density $\rho_0$, and predicts that the average density in a collapsing region grows with time as $\alpha$ flattens. This may indeed be the case for situations which allow for mass accretion onto GMCs. 
Mass accretion from the diffuse atomic envelop can provide a continuous supply of gas  which can increase the mean density.  However, many numerical box simulations use the condition of mass conservation and therefore the density shift can become important as material is funneled from the diffuse lognormal portion of the PDF to grow the powerlaw.

To instead renormalize the PDF to the average density inside a volume, a density shift ($s_s$) needs to be applied to obtain $s_{\rm{new}}$.  The density normalization condition is given as:
$\int_{-\infty}^{\infty} \rm{exp}(s)p(s)ds=1$
which results in the densities being shifted
$s_{\rm{new}}\leftarrow s-s_s$:
by
\begin{equation}
\label{eqn.shift}
s_s = \ln\left( \frac{\mathrm{e}^{s_t(1-\alpha)} NC}{\alpha-1}
+\frac{N}{2} \mathrm{erfc}\left( \frac{\sigma_s^2-2s_t}{2\sqrt{2}\sigma_s} \right) 
\right)
\end{equation}
{We refer the reader to \citet{Burkhart2018} for full definitions of the normalization constants N and C. }
We plot the shift density as a function of sonic Mach number (i.e. through the width of the PDF as given by Equation~\ref{eqn.sigma}) in Figure~\ref{fig:shift} for different values of PDF powerlaw slope ($\alpha$).
 The shift is small until $\alpha \approx 1.5$ and/or for low Mach numbers.  
 We will use the shift density when applying the model to the {\sc AREPO} simulations, which conserve mass.

\section{The Critical Density for Star formation and the Self-Gravitating Gas Fraction}
\label{sec:trans}

\subsection{The PDF Transition Density as a Critical Density for Star formation}

Here we provide a general overview of the post-shock critical density for gravitational collapse in a turbulent medium based on the presentation in \citet{krumholz05a} and \citet{Padoan11b} and its relation to the transition density given in \citet{Burkhart2017ApJ...834L...1B}.  We will show how this critical density, based on the post-shock density, can be related to the transition density from the lognormal to powerlaw portions of the density PDF. 

Supersonic turbulence in the ISM produces a cascade of energy that proceeds from parsec scales (or larger).
One consequence of this energy cascade is the line-width-size relation \citep{Larson81a,Solomon87a,Ossenkopf02a,Heyer2004,Wu10a,Wong11a,Barnes11a,Shetty11b,Hopkins12e,Kritsuk2013}, in which the turbulent velocity dispersion $\sigma_l$ computed
over a volume of characteristic length $l$ increases with $l$ as
$\sigma_l\propto l^{0.5}$ and can extend more than three orders of magnitude in length \citep{Ossenkopf02a}. As the velocity dispersion becomes increasingly damped as one goes to smaller scales within the turbulent cloud, there will be a scale at which the turbulence will transition from supersonic to subsonic, i.e., the sonic scale ($\ls$) \citep{Vazquez-Semadeni03a}. At this point, the linewidth size relation can shallow to the subsonic relationship: $\sigma_l\propto l^{1/3}$ \citep{Shetty12a,Chen2018}.

We define the sonic scale in a similar way as in \citet{Vazquez-Semadeni03a} and \citet{Krumholz2005}:
 let $\sigma_l$ be the one-dimensional velocity
dispersion computed over a sphere of diameter $l$ within a turbulent medium.  $\ls$ is defined as the length $l$ such that $\sigma_l=\cs$,
where $\cs$ is the isothermal sound speed in the region.

The linewidth-size relation can be normalized in terms of the sonic length:
\begin{equation}
\label{linewidth1}
\sigma_l=\cs \left(\frac{l}{\ls}\right)^p.
\end{equation}

For supersonic turbulence $p=0.5$ and we can then write the sonic scale as:
\begin{equation}
\label{linewidth2}
\ls= \left(\frac{L_{\rm cloud}}{M_s^2}\right)
\end{equation}

The other length scale of interest in a collapsing turbulent cloud is the Jeans length ($\lj$), which enters into the calculation of the classical  Bonnor-Ebert core, i.e.,  largest mass that an isothermal gas sphere embedded in a pressurized medium can have while still remaining in hydrostatic equilibrium \citep{Ebert55a,Bonnor56a}
\begin{eqnarray}
M_{\rm BE} & = & 1.18 \frac{\cs^3}{\sqrt{G^3 \rho}} \\
& = & \frac{1.18}{\pi^{3/2}} \rho \lj^3.
\end{eqnarray}

Where $\lj$ is:
\begin{equation}
\lj = \sqrt{\frac{\pi \cs^2}{G\rho}},
\end{equation}
The Jeans length is therefore the critical radius of a cloud where thermal energy is counteracted by gravity.

As discussed in \citet{Krumholz2005}, if $\lj \le \ls$,
gravity is approximately balanced by thermal plus turbulent pressure,
and the object is at best marginally stable against collapse\footnote{This will depend also on the strength of the magnetic pressure. Here we assume that the magnetic field is dynamically unimportant relative to turbulence and gravity. In other-words, the cloud is magnetically super-critical and super-Alfv\'enic.}. If
$\lj\gg\ls$, turbulent kinetic energy greatly exceeds both
gravitational potential energy and thermal energy, and the object is stable against collapse.
Since $\lj$ is a function of the local density, the condition $\lj
\le \ls$ for collapse translates into a minimum local density
required for collapse (in the absence of magnetic fields).  Equating the two length scales and solving yields a critical density:

\begin{equation}
\rho_{\rm crit}=\frac{\pi \cs^2 M_s^4}{G L_{\rm cloud}^2}
\end{equation}

We pause to consider some fiducial values for the critical density (in terms of number density):
\begin{equation}
n_{\rm crit}=1.3\times10^5 \left({\frac{\cs}{0.2 \rm kms^{-1}}}\right)^2\left({\frac{M_s}{10}}\right)^4\left({\frac{10pc}{L_{\rm cloud}}}\right)^2 \rm cm^{-3}
\end{equation}

For a sound speed of $\cs=0.2~{\rm km}~{\rm s}^{-1}$ and a range of $M_s=6$--$10$ and  L$_{\rm cloud}=5$--$10$ pc, we expect the critical density to be  $n_{\rm crit}\approx 2\times 10^4$--$1.3\times 10^5~{\rm cm}^{-3}$ over a length scale of $\lambda_s\sim\lambda_J=0.05$--$0.3~{\rm pc}$, which is the typical width of observed molecular cloud filaments \citep{Arzoumanian11a,Federrath2015,Panopoulou2017MNRAS.466.2529P}. These densities also provide a typical Bonnor-Ebert Mass of around a solar mass.

The critical density can be re-written in terms of the mean density, sonic Mach number and Virial parameter\footnote{The critical density derived in \citet{Padoan11b} has a similar form to that of \citet{Krumholz2005} but is derived  by comparing the Bonner-Ebert mass to the mass enclosed in a spherical region of density equal to the post shock density.}:

\begin{equation}
\label{eqn.s_c}
\rho_{\rm crit}/\rho_0=\rm{exp}(s_{\rm crit})=\frac{\pi^2}{15}\alpha_{\rm vir} M_s^2
\end{equation}
where constants included here arise from presuming a spherical volume with the radius being half the cloud length scale.
with $\alpha_{\rm vir}$  defined as:
\begin{equation}
\alpha_{\rm vir}=\frac{5v_L^2R}{GM}
\end{equation}

For molecular clouds, the typical virial parameters are low such that $\alpha_{\rm vir}$ is of order unity \citep{McKee03a,Kauffmann13a}. Therefore:
\begin{equation}
\label{eqn.critaprx}
\rho_{\rm crit}/\rho_0 \approx \rho_{\rm ps}/\rho_0 
\end{equation}

where the normalized post-shock density is:
\begin{equation}
\rho_{\rm ps}/\rho_0 \equiv M_s^2
\end{equation}
for an isothermal shock.

Thus for $\alpha_{\rm vir} \approx 1$ (i.e. for so-called virialized clouds)  the critical density is comparable to the hydrodynamic post-shock density $\rho_{\rm ps}$, within a factor of a few.
The physical meaning of this density is that the thermal pressure is comparable to the mean turbulent pressure in the cloud, $P_{\rm turb}$. Theses regions no longer experience enough thermal or turbulent pressure support to prevent collapse.

We can now relate the critical density for collapse to the transition density from the lognormal to powerlaw forms of the PDFs.
A PDF width-Mach number relation, e.g. Equation~\ref{eqn.sigma}, provides a direct link to a mathematically derived transition density and the physics of the critical density.  Combining Equation~\ref{eqn.sigma}, Equation~\ref{eqn.st} and Equation~\ref{eqn.s_c}:
\begin{align}
\label{eqn.trans_crit1}
s_t&=(\alpha -1/2)\rm{ln}\left(1+b^2\frac{15\rho_{\rm crit}} {\pi^2\alpha_{\rm vir}\rho_0} \right)
\end{align}

For the special case of  $\alpha=1.5$, the PDF transition density (Equation~\ref{eqn.st}) can be expressed as: 
\begin{align}
s_t&=\sigs^2
\end{align}

\begin{figure*}
\centering
\includegraphics[width=0.97\textwidth]{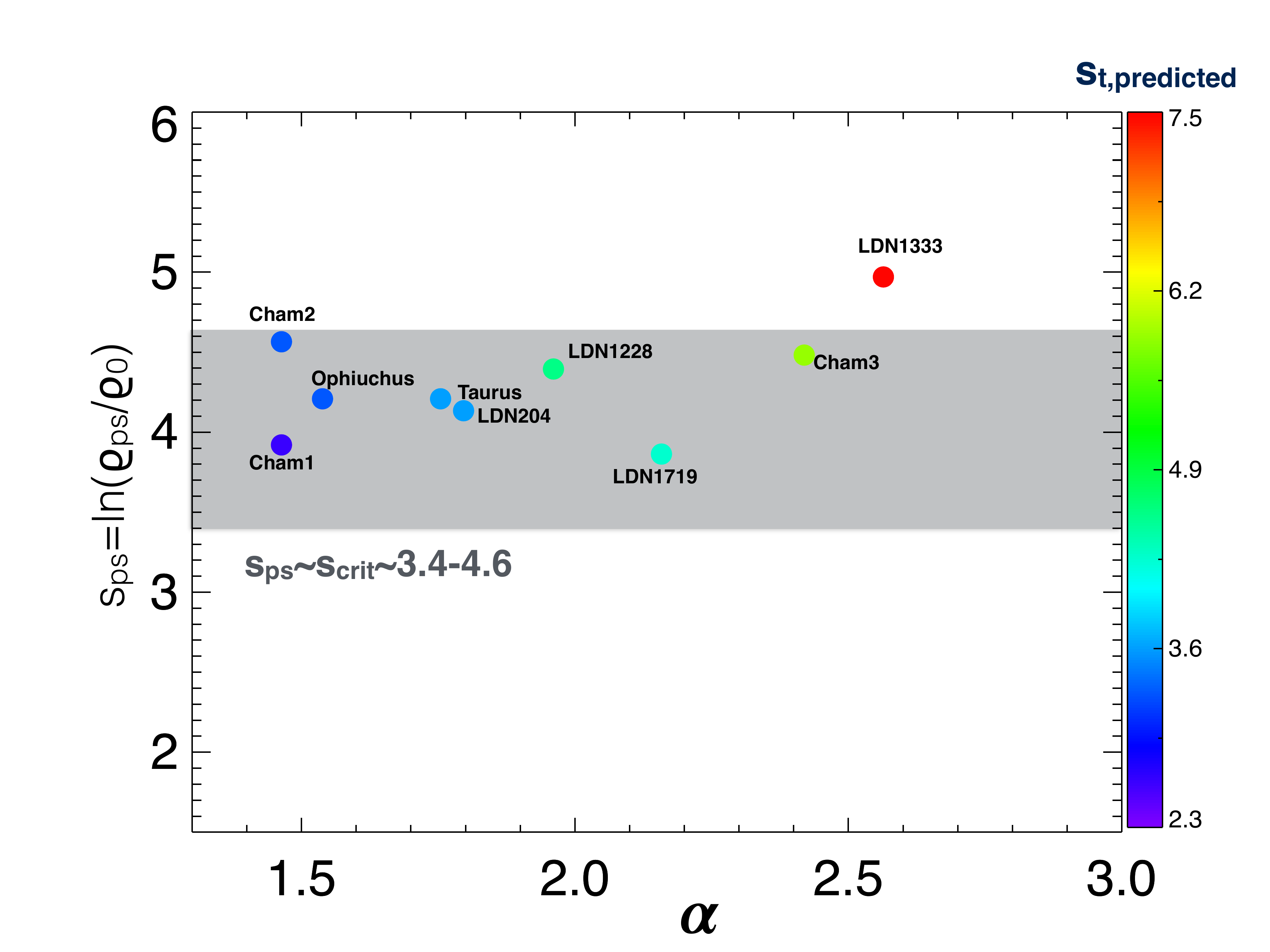}
\caption{
\label{fig:kain}
The post shock density vs. powerlaw slope for a number of observed GMCs.  The normalized post-shock density is derived from  Equation~\ref{eqn.critaprx} using values of the sonic Mach number and mean density from \citet{Kainulainen2014} and \citet{Kainulainen2017}. { \citet{Kainulainen2014} reported values of the radial density distribution power law slope ($\kappa$), which we use to calculate the density PDF powerlaw slope.} We find correspondence between the post shock density (y-axis) and the observational reported values of the critical density for collapse ($s_{\rm crit}\approx 3.4$--$4.6$, shade box) suggesting that the post-shock density is a reasonable indicator for collapse.  We compute the expected transition density for each cloud based on Equation~\ref{eqn.st}, as indicated in the color bar. $\sigma_s$ is based on the measured sonic Mach number and an assumed $b=0.5$, {choosing a $b$ value in between fully solenoidal and fully compressive.} As expected from Equation~\ref{eqn.trans_crit1}, $s_t\le s_{\rm ps}\approx s_{\rm crit}$ in the limit of $\alpha \approx 2$--$1.5$.  }
\end{figure*}

What is the observed critical density for collapse and how does it relate to Equation~\ref{eqn.trans_crit1}? The critical density was observationally determined by \cite{Lada10a} to be $n_{\rm crit}\approx 10^4$~cm$^{-3}$ and by \cite{Evans14a} to  be $n_{\rm crit}\approx 6.1 \pm 4.4\times 10^3$ cm$^{-3}$.  \citet{Kainulainen2014} found similar values for clouds on the verge of forming stars.   {In addition to measuring the critical density, \citet{Kainulainen2014} and \citet{Kainulainen2017} published values of the sonic Mach numbers, cloud mean densities, and PDF powerlaw tail slopes from measured values of radial density profile indices ($\alpha=3/\kappa$). } Using these values (e.g. Tables S1 and 1 of \citet{Kainulainen2014} and \citet{Kainulainen2017}), we calculate the expected post shock density  and predicted transition density $s_t$ (using the $b=0.5$, $\alpha_{\rm vir}=1$  and the observed sonic Mach number) to determine the PDF width\footnote{The PDF widths derived in \citep{Kainulainen2017} are overestimated in the context of the model presented here since they fit a single lognormal PDF rather than a lognormal + power-law PDF.} via Equation~\ref{eqn.sigma} and present the results in Figure~\ref{fig:kain}. We overplot a shaded horizontal box showing the approximate critical density derived from observational studies \citep{Lada10a,Evans14a,Kainulainen2014}.

\citet{Kainulainen2014} used YSO counts to determine an observational critical density of $s_{\rm crit}\approx 4$, which is in rough agreement with the prediction of Equation~\ref{eqn.critaprx} for clouds with $M_s\approx 7$--$14$.   In general, the post shock density shows good correspondence to the critical density, $s_{\rm ps}=3.4$--$4.6$ \citep{Lada10a,Kainulainen2014,Evans14a}, despite the unknown virial parameter and magnetic state of the clouds. Here we assumed  $\alpha_{\rm vir}=1$ however some clouds may be super-Virial \citep{Evans14a,padoan2017ApJ...840...48P}. More accurate measurements of the Virial parameter would be beneficial for future comparisons. 
If the clouds are super-Virial, the actual critical density would be higher than what we estimate here based solely on the measured sonic Mach number.  For this reason we label the y-axis in Figure~\ref{fig:kain} as the post-shock density. 
For shallow powerlaw slopes (x-axis values near $\alpha=1.9$--$1.3$, boxed in black) the PDF transition density (colored points) also shows good correspondence to the critical density and post-shock density, as expected from simulations \citep{burkhartcollinslaz2015,Mocz2017,Chen2017} and described by Equation~\ref{eqn.trans_crit1}. The relationship between MHD shock structure and the collapse is explored further in our companion paper, \citet{mocz2018MNRAS.480.3916M}

\subsection{The Self-Gravitating Gas Fraction}
\label{sec:dense}

In the last section we motivated the use of the PDF transition density as a critical density for star formation. This is further validated by recent observations of atomic gas and diffuse molecular gas in and around GMCs which has revealed that this diffuse material builds the lognormal form of the PDF while only dense molecular gas (traced by dust extinction/emission or HCN) resides in the powerlaw \citep{Burkhart2015,schneider2015MNRAS.453L..41S,Imara2016,Bialy2017ApJ...843...92B,Lombardi2015AA,Alves2017AA}.  
A study by \citet{Chen2018} verified that gas with low virial parameter resides primarily in the powerlaw portion of the gas PDF.

In what follows, we consider \textit{all the gas above the transition density (i.e., the gas in the powerlaw portion of the density PDF) to be dense ``self-gravitating gas'' and all the gas in the lognormal portion to be ``diffuse unbound molecular gas''.} The predicted self-gravitating gas fraction ($f_{\rm dense}$) can then be related to a star formation efficiency.  $f_{\rm dense}$ tells us the overall fraction of gas mass available for star formation at a given time based on the slope of the density PDF and the cloud environment (e.g., sonic Mach number).

We write the self-gravitating gas fraction as:

\begin{equation}
f_{\rm dense}\equiv \frac{M_{PL}}{M_{LN}+M_{PL}}
\end{equation}

or, more explicitly in terms of the density PDF:

\begin{equation}
\label{eqn.fdense0}
f_{\rm{dense}}=\frac{\mathlarger{\int}_{s_t}^{\infty}{\rm{exp(s)} P_{PL}(s) \deriv s}}{\mathlarger{\int}_{-\infty}^{s_t}{\rm{exp(s)}  P_{LN}(s) \deriv s}+\mathlarger{\int}_{s_t}^{\infty}{\rm{exp(s)} P_{PL}(s) \deriv s}}
\end{equation}

which evaluates to:

\begin{align}
\label{eqn.fdense1}
f_{\rm{dense}}=\frac{Ce^{(1-\alpha)s_t}}{\frac{\alpha-1}{2}\left(1+\rm{erf}\left(\frac{2s_t-\sigma_s^2}{2\sqrt{2}\sigma_s}\right)\right)+Ce^{(1-\alpha)s_t}}
\end{align}

Combining Equations~\ref{eqn.st} with~\ref{eqn.fdense1} we have:

\begin{align}
f_{\rm{dense}}(\alpha,\sigma_s)=\frac{Ce^{\sigma_s^2(1-\alpha)(\alpha-1/2)}}{\frac{\alpha-1}{2}\left(1+\rm{erf}\left(z\right)\right)+Ce^{\sigma_s^2(\alpha-1/2)(1-\alpha)}}
\end{align}

where $z=\frac{2\sigma_s^2(\alpha-1/2)-\sigma_s^2}{2\sqrt{2}\sigma_s}$

The fraction of gas in the powerlaw tail is therefore dependent only on the width of the lognormal and the slope of the powerlaw tail.  It is important to note that to derive the self-gravitating gas fraction relationship there is \textit{no need to invoke a critical density of collapse.}  The transition density between lognormal and powerlaw is determined solely by the properties of the density PDF itself, and therefore there are only two controlling parameters in the self-gravitating gas fraction.

\begin{figure*}
\centering
\includegraphics[width=0.97\textwidth]{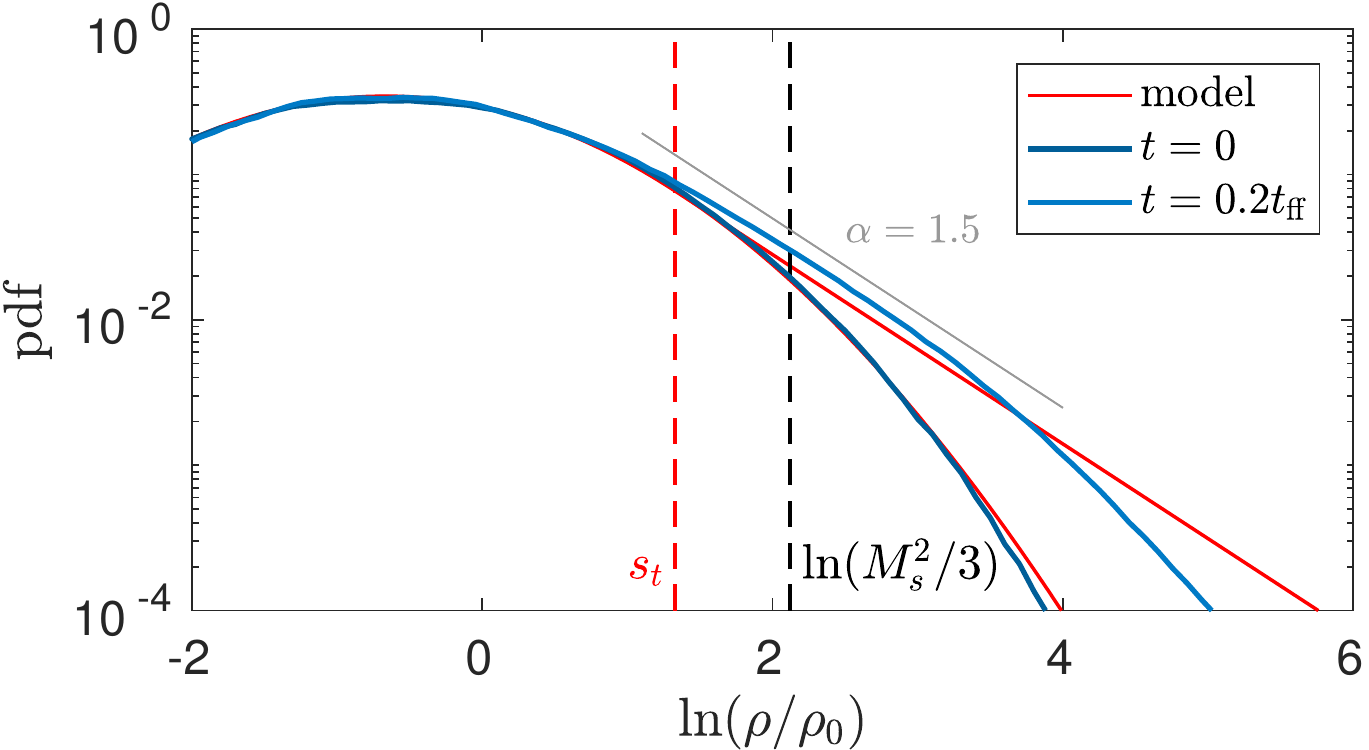}
\caption{
\label{fig:pdfExample}
Example of the gas density PDF from our {\sc AREPO} simulations (with $b=1/3$) that has a lognormal at t=0 (dark blue curve) and by $t=0.2t_{\rm ff}$ has generated an $\alpha=1.5$ powerlaw tail (light blue curve). The transition density ($s_t$) is identified in the $t=0.2t_{\rm ff}$ PDF.  We note that this simulation has no magnetic fields and therefore the powerlaw slope becomes flat significantly faster than MHD simulations discussed in previous works \citep{Collins12a,Burkhart2017ApJ...834L...1B}. The high-density end of the powerlaw takes longer to saturate but this does not strongly affect the value of $f_{\rm dense}$, as shown in Figure~\ref{fig:fdense}. For comparison, the critical density at which the background level of turbulent pressure is sub-dominant to the gas pressure ($\rho_0M_s^2/3$) is also shown and is related to $s_t$ analytically for $\alpha=1.5$.}
\end{figure*}

\subsection{The Critical Density for Collapse and PDF Transition Density in {\sc AREPO} Gravoturbulent Simulations}
\label{section.numerics}

We now test the above claims for relationship between the critical density for collapse and the PDF lognormal to powerlaw transition density with gravoturbulent simulations without magnetic fields. In particular, we use the {\sc AREPO} moving mesh code to perform numerical experiments similar to  \cite{Mocz2017}.
We identify the transition density and calculate the self-gravitating gas fraction over a free-fall time in {\sc AREPO} simulations of self-gravitating turbulence.  We drive solenoidal turbulence $(b=1/3)$ at sonic Mach numbers $M_s=5,10,16$, in the absence of magnetic fields, which generates the well-known lognormal density distribution. Self-gravity is then turned on, 
and the cloud is now also characterized by the virial parameter, with value $\alpha_{\rm vir}=1$ {and a box size of 5~pc.} Collapse leads to the development of a growing power-law tail in the density distribution. {The simulations are not designed to match a particular galaxy or GMC complex, but rather to test the analytic relationship for the critical density and transition density and the dense self-gravitating gas fraction. }

The transition density is identified in the resulting PDF as a function of time (i.e., as a function of $\alpha$) by fitting a lognormal and powerlaw function \citep{Burkhart2017ApJ...834L...1B}. 
{The fitted transition density reported here is identified robustly using an integral constraint as follows: (i) first we perform a linear fit to the PDF tail in log space to obtain the value of $\alpha$ in the simulations; (ii) then, we find the density in the simulations above which the dense gas fraction in the simulation matches that predicted by the analytic model of \citet{Burkhart2017ApJ...834L...1B}, given the measured $\alpha$. Only if the density PDF is smooth and continuous, and shows no additional features (e.g. strong deviations from non-lognormality), are the measured and predicted transition densities given $\alpha$ expected to be consistent.}

To compare the model with the simulation data, we also apply the shift in the normalization of the PDF (Equation~\ref{eqn.shift}) to renormalize the simulation PDFs. 
The shift density derived in Section 2 should be used in the case of mass conserving systems, e.g. for simulations with periodic boundary conditions.  We do not apply it later in the paper to observational data, as real GMCs are not isolated from their environment and may continue to accrete diffuse HI and H2 gas.  Figure~\ref{fig:pdfExample} shows an example from one of our numerical simulations with $M_s=5$ highlighting the fact that the gas density PDF is in fact continuous and that the transition density $s_t$ we have defined is in good agreement with the numerical simulations. 
For comparison, the critical density at which the background level of turbulent pressure is sub-dominant to the gas pressure ($\rho_0M_s^2/3$) 
\citep{Krumholz2005,LiPak2015} is also shown, which occurs above $s_t$, as predicted in Equation~\ref{eqn.trans_crit1}.

\begin{figure*}
\centering
\includegraphics[width=0.97\textwidth]{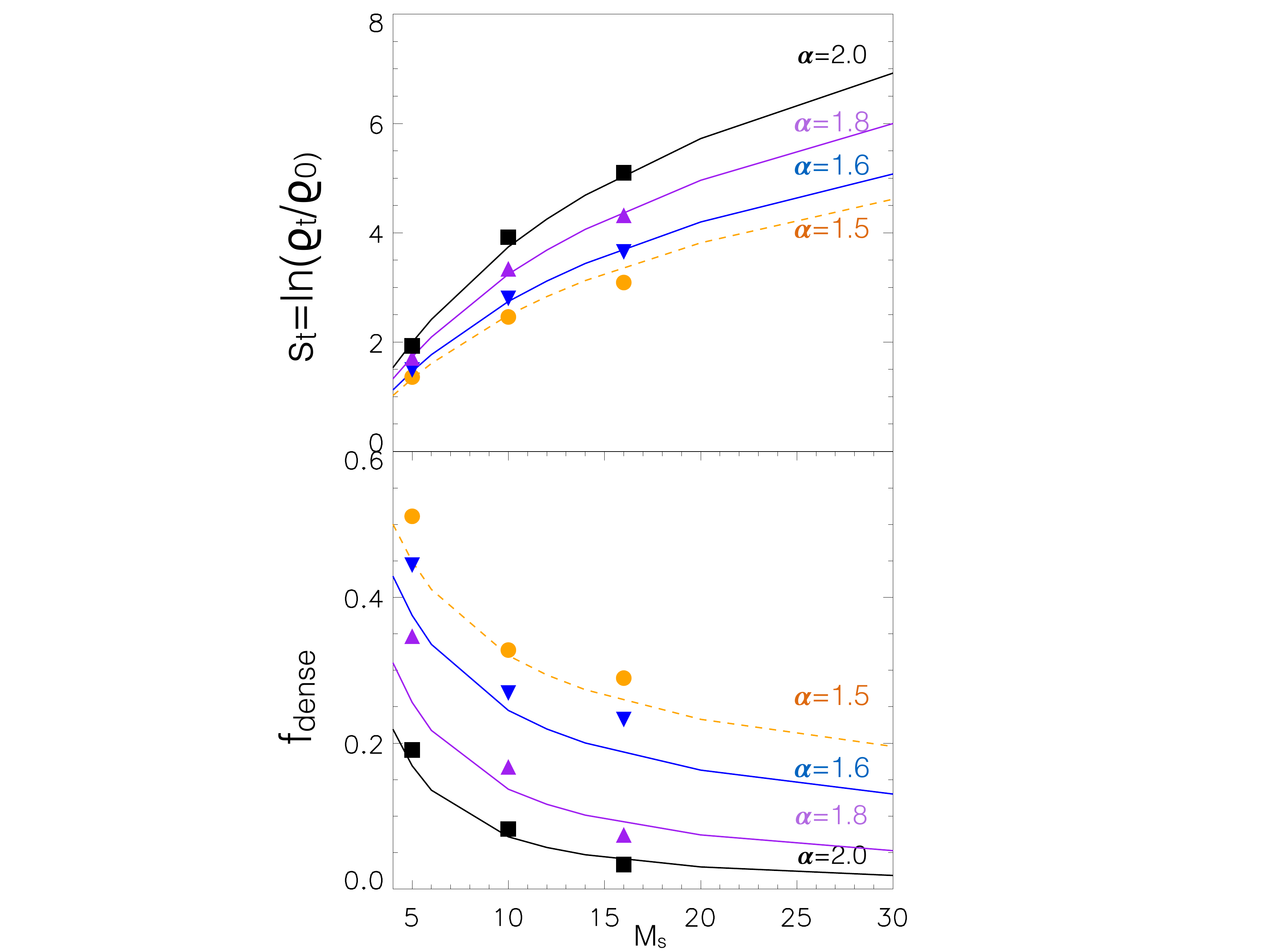}
\caption{
\label{fig:fdense}
Comparison of {\sc AREPO} simulations (individual points) to the theoretical model discussed here. Top panel: The transition density vs. sonic Mach number as given by Equation~\ref{eqn.st}.  As $\alpha$ becomes shallow the transition density moves towards lower densities which implies more dense gravitationally bound gas.  As the sonic Mach number becomes larger, the transition density moves to slightly higher values which implies slightly less dense gravitationally bound gas. Bottom panel: self-gravitating gas fraction vs. sonic Mach number, which reflects the trends of the transition density above. The simulations have $b=1/3$.}
\end{figure*}

{The observed transition density in the simulations are included as points in Figure~\ref{fig:fdense}, which match the analytic lines closely.  We also plot $f_{\rm dense}$ vs. $M_s$ in Figure~\ref{fig:fdense}, which is the predicted self-gravitating gas fraction as calculated by taking the fraction of gas above the transition density between lognormal and powerlaw. We find that the self-gravitating gas fraction is slightly anti-correlated with sonic Mach number.  This is because $s_t$ moves towards higher values with large sonic Mach number (PDF width) and there is less mass in the powerlaw portion of the PDF.  In the next section we will relate the self-gravitating gas fraction to the star formation efficiency. }

Our concluding result from this section is that the PDF transition density is analogous to the critical density of collapse \citep{Burkhart2017ApJ...834L...1B}.  The consequence of this relationship is that, for most actively star formation clouds with strong powerlaw tails, \textit{one may reduce the star formation rate calculation to a integral over the powerlaw material only}.

\section{The Star Formation Law Based on Self-Gravitating Gas Fraction}
\label{section.SFE}

{The self-gravitating gas fraction derived in the previous section represents the fraction of self-gravitating gas which is available for star formation. Some fraction of this gas will be expelled due to feedback, but assuming no other forces prevent gravitational collapse this represents the fraction of the GMC mass converted into stars in a transition density free fall time ($t_{\rm ff}(\rho_t)$). We can therefore theoretically relate the mass in the power law tail to the mass that will form into stars, modulated by some amount of material which is expelled from feedback.}

{To relate the power law PDF mass to the SFE we begin with some basic definitions.  A quantity of theoretical interest is the so-called integrated star formation efficiency defined as
the fraction of the gas mass that is converted to
stars across the entire lifetime of a cloud:}
\begin{equation}
\epsilon_{\rm int} \equiv \frac{M_{*}(t \rightarrow \infty)}{M_{\rm gas}(t=0)}
    \end{equation}

where $M_{*}(t \rightarrow \infty)$ is the total mass in stars formed at the end of the cloud's star forming life and $M_{gas}$(t=0) is the total initial gas mass, which has a density PDF in the form of a piecewise lognormal plus powerlaw. A quantity which is much more straightforward to measure observationally is the  “instantaneous” star formation efficiency: 
the mass fraction of stars
associated with the star-forming cloud at a given time,
\begin{equation}
\epsilon_{\rm inst}(t)  \equiv \frac{M_*(t)}{M_{\rm gas}(t)+M_*(t)}
    \end{equation}

$\epsilon_{\rm inst}$ will go from zero to some finite value during the star forming lifetime of a cloud and as $t\rightarrow\infty$, $\epsilon_{\rm inst}\rightarrow\epsilon_{\rm int}$. 

Finally, the star formation efficiency \textit{per free fall time} \citep{Krumholz2005,Lee2016} is given by:
\begin{equation}
\label{epsff}
\epsilon_{\rm ff}\equiv \epsilon_{\rm int}\frac{t_{\rm ff,0}}{t_*} 
\end{equation}
Where $t_*$ is the protostellar lifetime\footnote{The protostellar lifetime  varies from $t_*\approx 10^5$ years \citep{Mottram11} to several Myr \citep{Lee2016}.  A study of ten GMC complexes by \citet{Lee2016} yielded a ratio of free fall time to protostellar lifetime of $\frac{t_{\rm ff}}{t_{*}}=1.1$--$2.3$ (i.e. see their Table 3). } and the definition used here is the same as that used in \citet{Lee2016}.

We now connect these definitions of the star formation efficiency to the dense self-gravitating gas fraction, given by the ratio of the mass in the power law portion of the PDF to the total mass of the cloud.
The instantaneous stellar mass formed ($M_*(t)$) will be the mass in the powerlaw tail modified by some efficiency set by feedback, both of which are most certainly also functions of evolution in the cloud as the IMF evolves:  $M_*(t)=\epsilon_0(t)M_{PL}(t)$. 

\begin{figure*}
\centering
\includegraphics[width=0.97\textwidth]{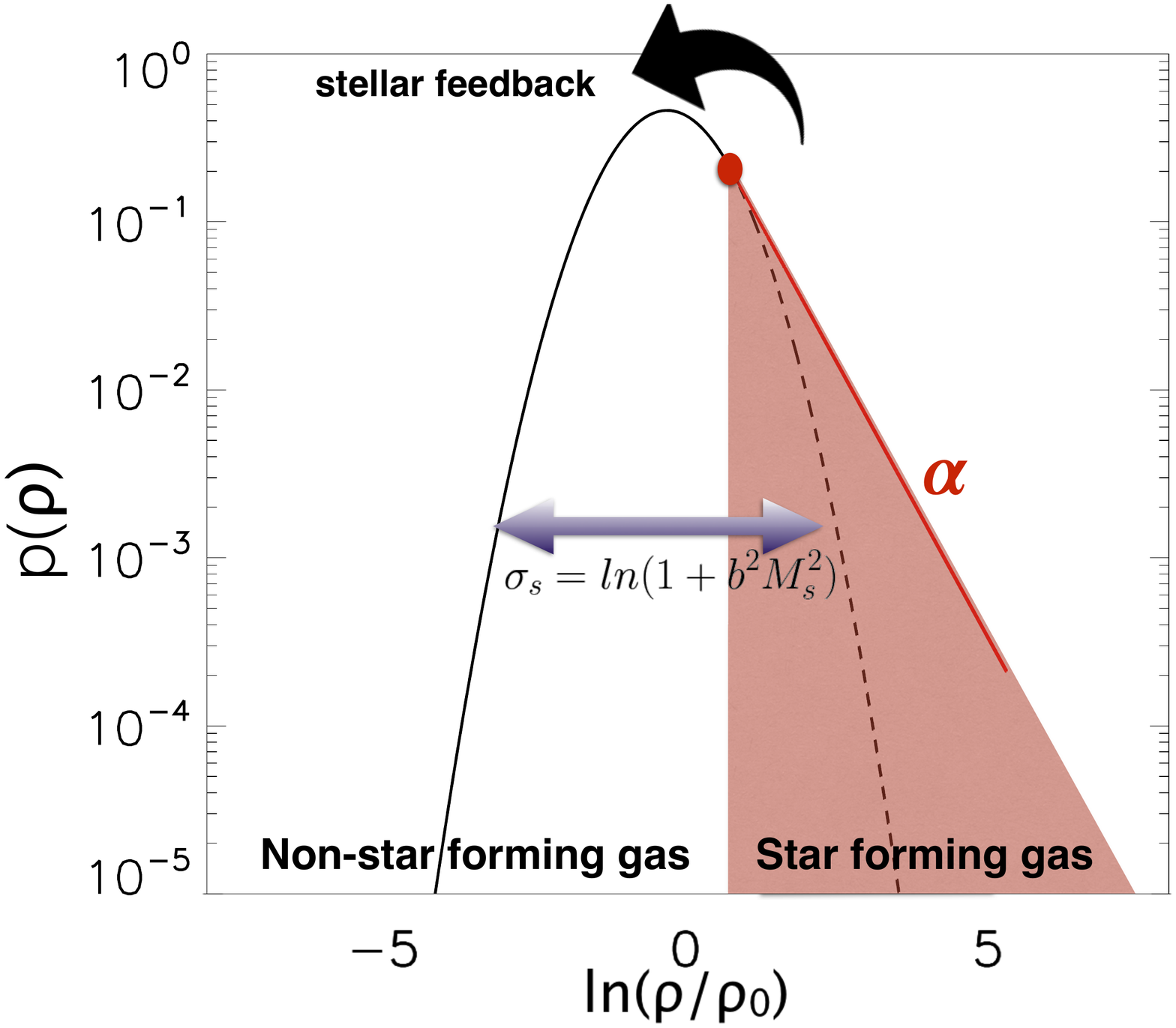}
\caption{
\label{fig:cartoonmodel}
Cartoon of model star formation efficiency using only the powerlaw tail to determine the fraction of self-gravitating gas. }
\end{figure*}

$\epsilon_0$ is a parameter that accounts for the fact that feedback will expel some fraction of the star forming self-gravitating gas; this will depend on the number of stars formed and their stellar type.   Feedback will expel gas from the power law tail and increase the diffuse lognormal portion of the PDF. Thus we expect the interplay of stellar feedback and gravity to regulate the shape of the density PDF and the overall ratio of the dense self-gravitating gas fraction in the cloud.
Figure~\ref{fig:cartoonmodel} shows a cartoon illustration of the density PDF of star forming and non-star forming gas.

We define the instantaneous star formation efficiency in terms of the density PDF as:

\begin{equation}
\label{eps}
\epsilon_{\rm inst}(t)=\frac{\epsilon_0(t)M_{PL}(t)}{M_{PL}(t)+M_{LN}(t)}=\epsilon_0(t)f_{\rm{dense}}(t)
\end{equation}

We note that $M_{PL}$ contains all mass with densities greater than the transition density, which includes the mass in both gas and stars, since the domain of integration in Equation~\ref{eqn.fdense0} goes to infinity.

{For the rest of the paper we will assume that $\epsilon_{\rm inst} \approx \epsilon_{\rm ff}$ however this is likely an overestimate \citep{Mottram11}.  We plan to investigate explicit time dependence in a future work, however we point to a recent numerical study by \citet{Grudic2018}, which studied how the integrated and free-fall efficiency varies with time in {\sc GIZMO} MHD gravoturbulent simulations with feedback (radiation, winds and supernova). They find that $\epsilon_{\rm ff}$ is on average, over the star forming life of the cloud, of the same order
as $\epsilon_{\rm inst}$, which is ∼ 100\% without stellar feedback and several
per cent with stellar feedback.  This suggests most of the stars form within the first cloud free fall time. }

{We especially note that there are two ways feedback can enter to the calculation of the SFE in Equation~\ref{epsff}.
One is through the factor $\epsilon_0$ and the other,  perhaps more physically meaningful approach, is that feedback self-regulates the overall ratio of self-gravitating to non-self-gravitating gas.  Feedback will expel gas from the dense, collapsing regions, increasing the diffuse portion of the PDF below the transition density  and steepening the powerlaw (see Figure~\ref{fig:cartoonmodel}). This will reduce the analytic calculation of the SFE since there is less dense self-gravitating gas in the powerlaw portion of the PDF (e.g. Figure~\ref{fig:fdense}, bottom panel).  Including a $\epsilon_0$ term was required of lognormal only models \citep{Krumholz2005,federrath12,Salim15a} as such models have no other way of introducing the effects of feedback, which are required to set SFE $\approx 1$\% \citep{Federrath2015,Semenov2017,Kruijssen2018}.  
We opt to keep $\epsilon_0=1$, for the rest of the paper to remove this parameter and focus on how the changing ratio of dense self-gravitating gas mass to total gas mass might match observations at different stages of cloud evolution and over the cloud's star forming lifetime. }

We pause to point out the difference between the calculation of $\epsilon_{\rm inst}$ derived from the dense self-gravitating gas fraction and the star formation rate per free fall calculated in Paper 1 \citep{Burkhart2018}. 
\citet{Burkhart2018} considered the initial stages of star formation in which the distribution of density transition from diffuse turbulent molecular gas to self-gravitating dense molecular gas.  This transition begins to occur once the gas reaches the aforementioned critical density for collapse ($s_{\rm crit}$, Equation~\ref{eqn.s_c}) and beings to form a powerlaw tail in the density PDF at high density.  The calculation of the star formation efficiency presented here is meaningful only when a powerlaw tail already exists, i.e. the cloud is already actively star forming and $s_t \approx s_c$.

The gas depletion time is given by:
\begin{equation}
\label{eqn.tdep}
\tau_{\rm depl} \equiv \frac{t_{\rm ff}}{\epsilon_{\rm ff}}
\end{equation}

\begin{figure}
\centering
\includegraphics[width=0.5\textwidth]{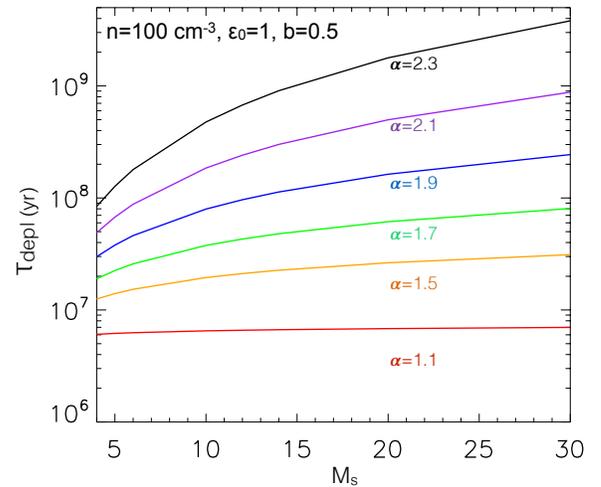}
\caption{
\label{fig:tdepl}
The depletion time calculated from the powerlaw slope PDF self-gravitating gas fraction model (Equation~\ref{eqn.tdep}) vs. sonic Mach number. We assume $\epsilon_0=1$ and vary the powerlaw slope ($\alpha$) to mimic the effects of self-regulated feedback. All curves are with $b=0.5$ and cloud mean free fall time of $t_{\rm ff}=4.3$~Myr (corresponding to a number density of 100cm$^{-3}$).  }
\end{figure}

We plot the depletion time in Figure~\ref{fig:tdepl} for a mean cloud free fall time of 4.3~Myr (corresponding to a mean number density of $n=100~{\rm cm}^{-3}$). In Figure~\ref{fig:tdepl}  we choose to leave $\epsilon_0=1$ and vary only $\alpha$.   We find reasonable agreement with the observed value of 1--2~Gyr with $\alpha > 2$ without needing to modify the feedback parameter.  Typical depletion times in extragalactic CO observations are around 1--2~Gyr \citep{Wong2002,leroy2008,Bigiel2008,Schruba2011,Bigiel2011,Leroy13a,Leroy2017} while dense gas tracers (such as HCN) can have depletion times as short as 20~Myr \citep{Gao04a,Lada12a}. This diversity in depletion time is reproduced for our range of powerlaw slope values.  {If the free fall time is increased, as might be applicable to beam diluted observations of molecular gas, the depletion time should also increase and the curves will move up linearly. Similarly, if $b$ or the Mach number is increased, the depletion time increases.}
Steep powerlaws ($\alpha=2$) may correspond to a higher fraction of diffuse molecular gas.  This diffuse gas can be a significant filling fraction in extragalactic observations which have larger beam sizes \citep{Kruijssen2018}.  Our model predicts the longest depletion times when the powerlaw slope is steep.  In the case of more dense self-gravitating gas, $\alpha$ shallows and the depletion time decrease to HCN traced values in Galactic clouds of several hundred mega-years.

\subsection{Comparison to Observational Datasets}
\label{section.obs}

Perhaps the most interesting effect of  Equation~\ref{eqn.tdep} is the prediction that \textit{the depletion time should slightly increase with increasing sonic Mach number}.  The increase of depletion time with velocity dispersion has been observed recently in M51 from the PdBI Arcsecond Whirlpool Survey (PAWS, \citet{Schinnerer13a,Leroy2017}). This suggests that higher sonic Mach number (higher velocity dispersion) induces increased support in the diffuse gas. 

In addition to the depletion time we can use the self-gravitating gas fraction to compute the star formation law:

\begin{equation}
\Sigma_{\rm SFR}=\epsilon_{\rm ff}\frac{\Sigma_{\rm gas}}{t_{\rm ff}}
\end{equation}

and can compare the model trends with sonic Mach number with recent observations.

We compare the  model predictions with published data from \citet{Leroy2017} on M51 from the PdBI Arcsecond Whirlpool Survey (PAWS, \citet{Schinnerer13a}). 
PAWS mapped CO (1-0) emission from the inner $9\times6$~kpc of M51. At the PAWS resolution of 40~pc, the structure of the turbulent
ISM at the scale of an individual giant molecular
cloud \citep{Hughes13a,Colombo14a} can be partially resolved. 
Combining this information with infrared maps
from Herschel and Spitzer \citep{Kennicutt03a}, \citet{Leroy2017} measured how the cloud-scale structure of the ISM relates to M51's ability to form stars.  They found that the star formation efficiency correlates strongly with the strength of self-gravity, in agreement with our model.  The velocity dispersion is found to anti-correlate with SFE, in contrast to  lognormal turbulence regulated star formation theories.

\begin{figure*}
\centering
\includegraphics[width=0.97\textwidth]{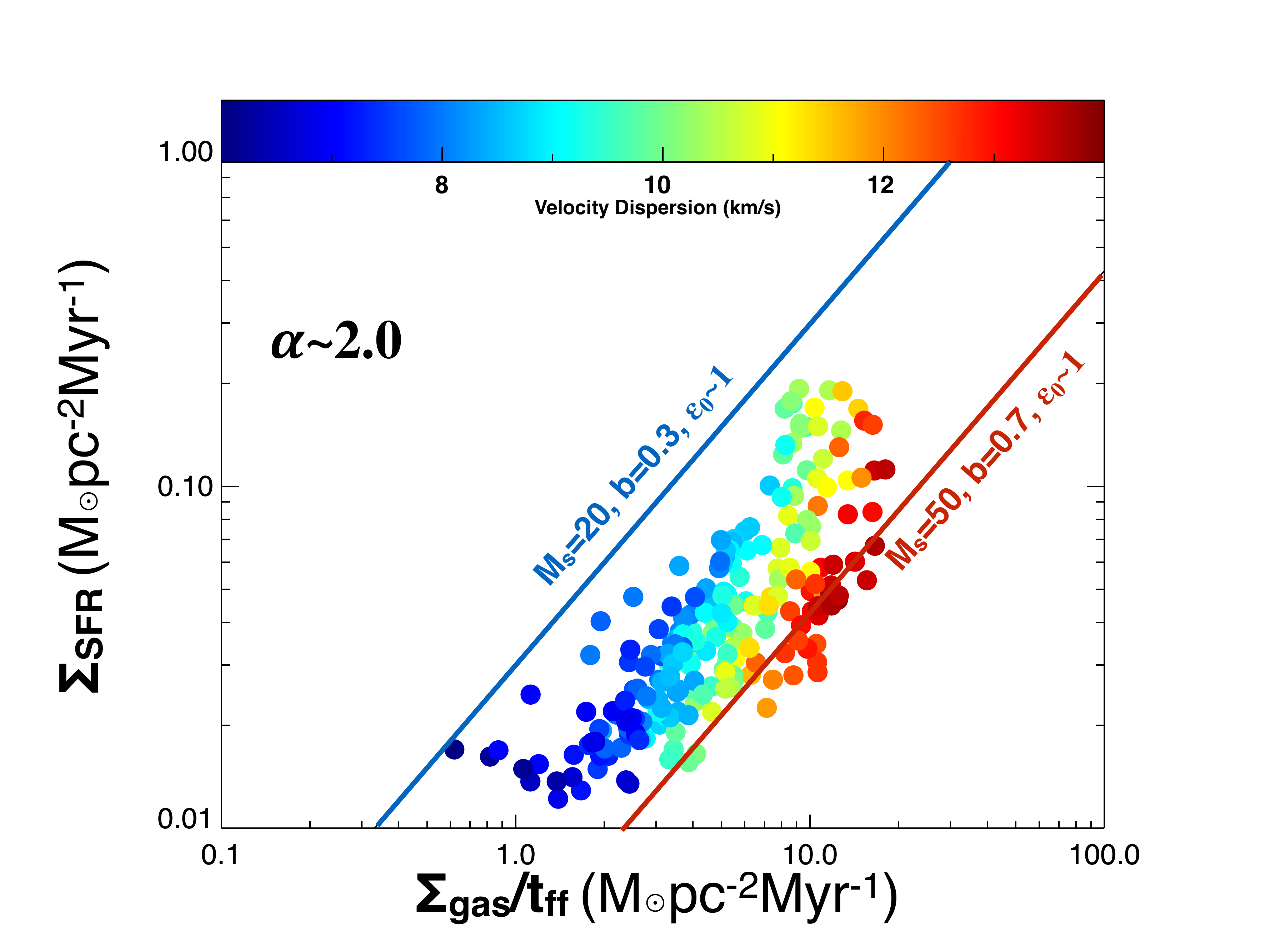}
\caption{
\label{fig:paws1}
The PAWS M51 star formation rate surface density vs. the gas surface density divided by the free fall time. The data are taken from \citet{Leroy2017} within a 30'' beam. {Different colors indicate the velocity dispersion measured in the individual 3~arcsecond regions in M51.} A clear gradient can be observed with larger rms velocities (larger sonic Mach number).  The blue and red lines correspond to our analytic model.  The blue line corresponds to $M_s=20$, $b=0.3$, and $\epsilon_0=1$. The red line corresponds to $M_s=50$, $b=0.7$. Both lines use $\alpha=2$. Larger PDF width (controlled by higher Mach number/compression) corresponds to lower  $\epsilon_{\rm ff}$ in agreement with the PAWS data. We do not need to use the efficiency factor to match the observations with $\alpha=2$.  }
\end{figure*}

We plot the PAWS M51 star formation rate surface density vs. the gas surface density divided by the free fall time in Figure~\ref{fig:paws1}. Each point represents an individual 30'' beam averaged regions in M51 (after beam averaging each point has a resolution of 1.1~kpc) and is color-coded with its corresponding velocity dispersion.
A clear velocity gradient can be observed with larger velocity dispersions towards larger values of $\Sigma_{\rm gas}/t_{\rm ff}$.  For a constant $\Sigma_{\rm SFR}$ this indicates lower SFE. 

We over plot lines of constant $\epsilon$ from {Equation~\ref{eps} }for a lognormal and powerlaw PDF model with $\alpha=2$ and include two different PDF widths (blue and red lines) which correspond to the range of $M_s$ and $b$ values possible from the CO observations. The blue line corresponds to $M_s=20, b=0.3$.  The red line corresponds to wider PDFs ($M_s=50$, $b=0.7$).  Larger PDF width (higher Mach number) corresponds to lower  $\epsilon_{\rm ff}$, in contrast to lognormal-only turbulence SFR theories and in agreement with the PAWS data.  This suggests that including the powerlaw tail in analytic star formation calculations is important. We choose to keep $\alpha=2$ for two reasons: first this allows us to match the observations without the need to invoke the fudge-factor $\epsilon_0$. Second the PAWS data are CO observations which are mostly tracing diffuse molecular gas and have large beam filling fractions with the 1.1~kpc resolution. As we expect diffuse gas to mostly be dominated by turbulence and not gravity, this would be suggestive of a steep powerlaw slope. We would expect dense gas tracers to be consistent with shallower values of $\alpha$ and hence larger dense (self-gravitating) gas fractions.

The PAWS data allow for constraints of the sonic Mach number between roughly 20 and 50 given a typical CO sound speed of $c_s=0.3$~km~s$^{-1}$.
This leave the only fully free parameters in the model the compressibility of the turbulence (b) and the powerlaw slope. In regards to the feedback efficiency, this parameter is highly uncertain and could depend on individual star forming environments (e.g. strong or weak magnetic field, \citet{Hull2017,Mocz2017}) and the stellar IMF. Feedback can also result in a steeper density PDF powerlaw slope as gas is expelled from dense star forming regions.   

It is encouraging that the CO observations can be reproduced without the $\epsilon_0$ parameter and a $\alpha=2$. To further test this claim, we apply a Monte Carlo approach to { Equations~\ref{eqn.fdense1} and~\ref{eps}} choosing a velocity dispersion range which matches the PAWS data (v$_{\rm rms}\approx 6$--$16$~km~s$^{-1}$) and a sound speed of $c_{s,{\rm CO}}=0.3$~km~s$^{-1}$. {We sample our Monte Carlo experiment using 300 draws and use a Mach number range of $20$--$53$. We allow for values of $\alpha=2.2$--$1.5$, $b=0.3$--$0.5$, and fix  $\epsilon_0=1$ to remove this free parameter. The idea being that the shape of the PDF alone will regulate the overall efficiency and a fudge factor should not be needed.} 

\begin{figure*}
\centering
\includegraphics[width=0.97\textwidth]{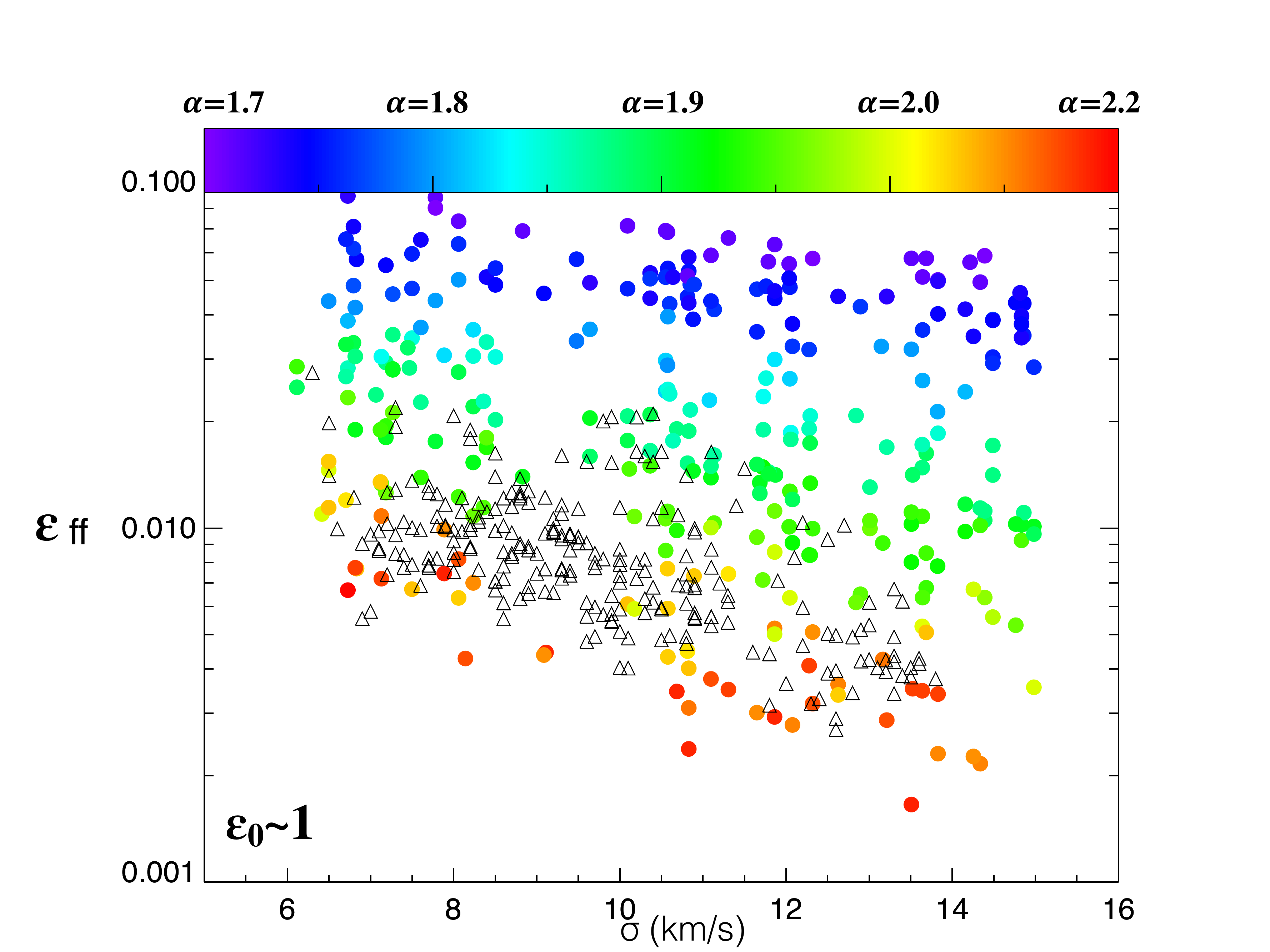}
\caption{
\label{fig:paws2}
$\epsilon_{\rm ff}$ vs. velocity dispersion for the PAWS M51 data (black diamonds) and for the analytic model Monte-Carlo data points (colored points, colored coded by values of $\alpha$).  
The data from the analytic model is able to reproduce the M51 CO star formation efficiencies  $\alpha=1.9$--$2.2$ and $M_s=20$--$50$, assuming a sound speed of $0.3~{\rm km}~{\rm s}^{-1}$. We do not need to use a feedback efficiency fudge factor (i.e. we set $\epsilon_0=1$).}
\end{figure*}

We plot the PAWS $\epsilon_{\rm ff}$ vs. velocity dispersion ($\sigma$) in Figure~\ref{fig:paws2}. The PAWS M51 data is shown in black diamonds and the analytic model Monte-Carlo data points are colored coded by values of $\alpha$.    We find excellent agreement with the data with  $\alpha=1.9$--$2.2$ and $M_s=20$--$50$. $\alpha<1.8$ over-predicts the PAWS $\epsilon_{\rm ff}$ and an efficiency factor would need to be used to lower the value to match, as is often done with lognormal PDF turbulence regulated star formation models.  If $\alpha=1.5$ the $\epsilon_0$ values are constrained between $2$--$8$\% in order to match the PAWS data range.  It is encouraging that the analytic model  allows us to remove the need for the $\epsilon_0$ factor and use the powerlaw tail as an indicator of the self-gravitating gas fraction and evolutionary state.  

\section{Discussion and Summary}
\label{section.sum}

In this section we discuss some interesting implications of our results. This includes connecting the model to observations, the dynamic SFE, the critical density for collapse and the hitherto neglected role of magnetic fields. 

\subsection{Connecting Theory and Observation}

The star formation efficiency presented in this work is computed from the dense self-gravitating gas fraction with free parameters being the slope of the powerlaw tail ($\alpha$), the width of the lognormal ($\sigma_s$). These two parameters set the self-gravitating gas fraction, which is {dynamic in time} given the complicated interplay of stellar feedback, gravity and turbulence. Other analytic star formation models are parameterized with a critical density, which in turn depends on parameters such as the Virial parameter, feedback efficiency, turbulence driving scale, Mach numbers, mean density and temperature, as well as other fudge-factors of order unity. With a number of free parameters to deal with, observational studies have had difficulty in confirming or disproving such theories. 

Observers can directly test the model presented here based on measurements of  $\epsilon_{\rm ff}$ and/or the self-gravitating gas fraction and the PDF powerlaw slope or transition density.  Direct measurements of the density PDF width from observations are complicated by line-of-sight-effects \citep{Lombardi2015AA,Chen2018}. Therefore, the connection to the model can be made via measurements of the sonic Mach number (velocity dispersions) and powerlaw PDF slope. For extragalactic observations, such as the PAWS M51 data discussed here, the powerlaw slope is not directly measured.  However we were still able to test the basic scalings of this model using CO velocity dispersion measurements and measured 
$\epsilon_{\rm ff}$.  {Another uncertainty with comparison the model to observations is the unconstrained nature of the forcing parameter b, which delineates between solenoidal ($b=1/3$) and compressive turbulence driving ($b=1$). While the type of forcing is difficult to constrain observationally, recent works have developed promising methods for quantifying the modes involved in producing turbulence \citep{herron2017MNRAS.466.2272H,kainfed2017A&A...608L...3K}.}

Is the density PDF model developed here applicable to galaxy scale observations, in which the beam likely encompasses many clouds?  In fact, PDF models for star formation have been applied to galaxy scale observations since the seminal paper of \citet{Krumholz2005}, which used a lognormal PDF model to predict the Kennicutt-Schmidt relationship. \citet{Federrath2017} applied lognormal PDF models to the SAMI galaxy survey to infer the gas surface density from their predicted SFRs. The PDF of HI and molecular gas tracers has been found to be lognormal and/or lognormal + powerlaw in extragalactic observations \citep{burkhart10,Maier2016AJ....152..134M,corbelli2018}.
 In our study, as well as in these example previous studies, the PDF sampled in the beam is averaged over many clouds in the same way that the SFR or other quantities of interest are beam averaged and therefore the comparison is meaningful as long as the scales traced are within the turbulence inertial range.  Outside of the turbulence inertial range, the PDF width no longer is related to the sonic Mach number. { Although the outer-scale of turbulence can be difficult to measure \citep{Pingel2018} we note that the largest scale drivers such as supernova and gravitational instabilities should operate on kiloparsec scales \citep{Stanimirovic2001,Shetty08a,Krumholz10c,krumholz2018MNRAS.477.2716K}. Additionally, there is observational evidence for kpc driving in dwarf galaxies, such as the Small Magellanic Cloud, from measured velocity power spectrum \citep{Chepurnov2015}.  Therefore changes in the PDF on scales of GMCs are more likely due to phase or equation of state changes \citep{Federrath2015} or the atomic to molecular transition \citep{Bialy2017ApJ...843...92B} rather than the lack of a turbulence cascade.}

\subsection{The Depletion Time and the Dynamic SFE}

Gas cycles in and out of the power-law portion of the PDF (i.e. the portion of the cloud which is collapsing and thus available for star formation)  as a result of stellar feedback and accretion.
The cycling of gas from dense collapsing regions to diffuse expanding/supported regions is, in part, why the depletion time of gas in galaxies is long compared to other relevant galactic dynamical time scales.  The simple calculation of the self-gravitating gas fraction presented here is able to reproduce observed values of the depletion time. 
The average depletion time for any given region will depend on the properties of the feedback \citep{Semenov2017,Grudic2018}. Future studies will constrain the time dependency of $\alpha$ with stellar feedback which may allow for the complete removal of $\epsilon_0$ from this model.  In our current study, observed extragalactic depletion times (i.e. of $1$--$2$~Gyr) can be reproduced with no fudge factors with $\alpha=2$, which may represent the integrated average value of the powerlaw slope over the cloud's star forming lifetime. The depletion time depends primarily on the amount of self-gravitating gas and secondarily on the turbulent environment \citet{Leroy2017}.

Observational and numerical estimates of the integral star-formation efficiency are in the range of $\epsilon=0.0001-0.2$ \citep{Evans09a,Lada10a,Ostriker11a,Krumholz12a,Zamora2014ApJ...793...84Z,Krumholz14c,Lee2016,Semenov2017,Grudic2018} with an average value of around 1\% \citep{krumreview2014}. Many of these simulations and observations also suggest that $\epsilon_{\rm ff}$ and $\epsilon_{\rm int}$ are \textit{not constant}, which is predicted by many lognormal theories of turbulence regulated star formation.  The model suggested here and in Paper 1 (Burkhart 2018) is that $\epsilon_{\rm ff}$ and $\epsilon_{\rm int}$ is dynamic on the scales of GMCs,  in agreement with data presented in \citet{Lee2016} and recent simulations with feedback, such as those of \citep{Grudic2018}. The self-gravitating gas fraction and the star formation rate is inherently dynamic if calculated from a density PDF whose powerlaw slope changes over the cloud evolution.  Initially the density powerlaw slope should shallow but once feedback becomes dynamically important it can steepen again.   The PDF transition density from lognormal to powerlaw presents a natural bisection between collapsing gas and supported gas. Near the transition, dense gas may be accelerated into free-falling regions.  Local accretion of gas will at some point be in competition with stellar feedback processes and ionization from more massive stars.

\subsection{The Meaning of the Critical Density for Collapse}

The post-shock density is an often invoked critical density which appears in several theories of star formation \citep{Krumholz2005,Padoan11b,federrath12}.
The relationship between the transition density and post shock density provides a natural critical density motivated by the effects of turbulence and gravity on the PDF.  In the star formation rate formulation here, we did not need to invoke any specified critical density as the conditions of continuity and differentiability of the lognormal plus powerlaw PDF provide an explicit definition of the transition density. The transition density is related to the post-shock density in the presence of a powerlaw PDF distribution. 

In general, for clouds on the verge of or just beginning to collapse and form a powerlaw tail (e.g. starless cores),  a critical density  residing in the lognormal is appropriate.  This was the approach taken in Paper 1 \citep{Burkhart2018} which calculated the SFR directly over the lognormal from the critical density and then the powerlaw from the transition density. However once the collapse proceeds, clouds form a powerlaw tail with shallow slope within a free fall time or less \citep{burkhartcollinslaz2015}. \citet{Burkhart2018} found that gravity dominates the SFR and SFE when the powerlaw is flat. At this evolutionary stage the SFR and SFE has little memory of the initial cloud turbulence.

\subsection{Effects of Magnetic Field on the Critical Density-PDF Transition Density Relationship}

We have thus far ignored any scenario in which magnetic fields strongly affect the PDF and the critical density for collapse.  In this subsection we address the possible effects of magnetic fields on our results.  We summarize the important physical parameters for magnetic fields in star formation and how these parameters can affect the critical density for collapse and PDF.  We will study the full effects of a strong magnetic field on the PDF and initial collapse in a future work.

There are three primary parameters which can encapsulate the importance of magnetic fields in the star forming interstellar medium on scales of clouds and pre-stellar cores (e.g. tens of parsecs to 100s AU scales):
\begin{itemize}
\item The Alfv\'enic Mach number: the ratio of the turbulent kinetic energy to the magnetic energy, which is defined as $M_A=\frac{V}{V_A}$, where $V$ is the turbulent velocity and $V_A=\frac{|B|}{\sqrt{4\pi\rho}}$ is the Alfv\'en speed. 
\item The plasma $\beta$: the ratio of the thermal pressure ($P_{\rm thermal}$) to magnetic pressure ($P_{\rm mag}$), which is defined as $\beta=\frac{P_{\rm thermal}}{P_{\rm mag}}$.  The Plasma $\beta$ can also be defined as $\beta=\frac{2M_A^2}{M_s^2}$

\item The mass-to-flux ratio: encompasses the importance of the magnetic energy to the gravitational potential energy.
This can be expressed in terms of the ratio of the cloud mass to the magnetic critical
mass, $M_\Phi$, which is the minimum mass that can undergo gravitational collapse in  a magnetically dominated medium.
In terms of the magnetic flux,
$\Phi\equiv B c_A \ell_0^2$,
the magnetic critical mass is
\begin{equation}
M_\Phi=c_\Phi\frac{\Phi}{G^{1/2}}\ ,
\label{eq:mphi}
\end{equation}
where $c_\Phi \approx 0.12$ for a cloud with a flux-to-mass distribution corresponding to a uniform field threading a uniform spherical cloud \citep{Mous76}.
For $c_\Phi=1/2\pi$, the ratio of the mass to the magnetic critical mass is
\begin{equation}
\mu_{\Phi,\,0}\equiv\frac{ M_0}{M_\Phi} \propto \frac{M_A}{\alpha_{\rm vir}^{1/2}}
\end{equation}
Which relates the virial parameter ($\alpha_{\rm vir}$) to the Alfv\'enic Mach number and magnetic critical mass.
The ratio
$\mu_{\Phi,0}$ is sometimes written as the ratio of the observed
mass-to-flux ratio to the critical one, $(M/\Phi)_{\rm obs}/(M/\Phi)_{\rm crit}$ 
(e.g., \citep{Troland08a,Crutcher2009,McKee2010,Crutcher12a}).

Gravitationally bound clouds that are both
magnetized and turbulent have $\mu_{\Phi,\,0}$ somewhat
greater than unity (i.e. are \textit{super-critical}) since the gravity has to overcome both the turbulent motions and the magnetic field \citep{McKee89a,LEC12}. If the cloud is \textit{sub-critical}, than the cloud can not collapse under ideal MHD due to the frozen-in condition and therefore a diffusion effect must be invoked, e.g., ambipolar diffusion or reconnection diffusion \citep{Zweibel1983,Mck10,LEC12}.

\end{itemize}

Additional important effects of magnetic fields, such as ion-neutral decoupling, depend on the above parameters in addition to the ionization fraction \citep{Balsara2010,Burkhart2015}.

The relationship between the PDF transition density and the critical density derived in the previous sub-section assumed that the cloud is super-critical, super-Alfv\'enic and has a $\beta>1$.  These conditions may be sufficient for most local GMCs \citep{Crutcher2009} and would explain the excellent correspondence between the post-shock density and the critical density in Figure~\ref{fig:kain}.  The correspondence between the post-shock density, PDF transition density and critical density for collapse has been reported in trans-Alfv\'enic and super-Alfv\'enic simulations \citep{Padoan11b,Collins12a,burkhartcollinslaz2015,Mocz2017,padoan2017ApJ...840...48P}.

The magnetic field can slightly alter the relationship between the PDF width and sonic Mach number.  For example, \citet{Molina2012} derived a formula that included the plasma beta: $\sigma_s={\rm ln}(1+b^2M_s^2\frac{\beta}{1+\beta})$.
However this relationship is valid only in the case of super-Alfv\'enic turbulence \citep{Molina2012}.  

We point out that the critical density for collapse may be controlled by very different physics for clouds which are  sub-critical and/or sub-Alfv\'enic. This is because the shock profiles can look very differently under the conditions of strong magnetic fields. We study this effect on the critical density for collapse further in a companion paper (Mocz \& Burhart 2018). Furthermore, for sub-critical mass to flux values, magnetic reconnection diffusion  will set a different critical density and critical length scale as discussed in \citet{LEC12}.   As shown in \citet{Mocz2017}, the transition density/critical density is not the same as the post-shock density when the medium is sub-critical and sub-Alfv\'enic \citep{LEC12}.  

Many atomic clouds (i.e. traced by the 21-cm emission line) have been found to be magnetically sub-critical while molecular clouds are in a super-critical state \citep{Crutcher12a}.  The transition from atomic to molecular media may also involve a transition from sub- to super-critical, which should involved a diffusion process. Most (or perhaps all) molecular clouds are found to be super-critical \citep{TroC08,Crutcher2009,Crutcher12a} and therefore we focused this paper on the critical density and density PDF without the inclusion of magnetic field effects (i.e. the super-Alfv\'enic/super-critical case).

\subsection{Additional Implications}

Our results also imply that molecular hydrogen is not required for star formation. At low redshifts, molecular hydrogen dominates the mass budget over atomic hydrogen at densities which are unable to gravitational collapse in GMCs. In otherwords, atomic gas transitions to molecular gas at lower densities than the critical density for collapse. This is naturally explained by 21-cm observations in the local universe where atomic gas demonstrates lognormal PDFs and other higher-order statistics of supersonic turbulence without signs of collapse \citep{Burkhart09,burkhart10,Zhang2012ApJ...754...29Z,burkhart15,Pingel2013,Imara2016,Maier2016AJ....152..134M,Maier2017AJ....153..163M,Nestingen2017ApJ...845...53N,Bialy2017ApJ...843...92B,Pingel2018}. Furthermore, dwarf galaxies with essentially no star formation and no molecular gas are observed in the extreme environment of galaxy clusters \citep{taylor2012MNRAS.423..787T,Janowiecki15a,Cannon15a,Burkhart2016,bellazzini2018MNRAS.476.4565B} again confirming that HI is unbound.   Our results suggest that even a large fraction of the molecular ISM is not collapsing or ultimately does not participate in star formation \citep{Chen2018}.
If the transition density for H2 formation is shifted to higher densities, as is probably the case for extreme low metalicity systems, star formation could proceed in atomic gas \citep{Krumholz12e} and atomic gas could develop a powerlaw PDF. The atomic gas phase would be able to reach higher density before self-shielding to form H2. 

Finally, the relationship between the PDF transition density and the post-shock density provides a natural explanation for why star formation proceeds in filamentary substructures.  Many filament models suggest the link between self-gravitating supersonic turbulence and filamentary structure in star formation. Filaments in GMCs have been shown to be the precursors to sub-Virial cores which make up the powerlaw portion of the density or column density PDF, as is seen in both simulations and observations \citep{Chen2018}. In the absence of self-gravity, the post-shock density and sonic scale provide a typical filament scale and density \citep{Federrath2015}. { If the additional condition 
of $\lambda_J \le \ls$ is met these post-shock filaments can become self-gravitating and produce a powerlaw radial density profile, e.g., $\rho\propto r^{-\kappa}$ \citep{shu77,Kainulainen2014,MyersP2015,MyersP2017}.    }

\section{Conclusions}
\label{sec:con}
 We use the analytic model of Burkhart (2018) to calculate the star formation efficiency and self-gravitating gas fraction in the presence of self-gravitating super- or trans-Alfv\'enic turbulence using a piecewise lognormal \textit{and} powerlaw density PDF. 
In summary we find that:

\begin{itemize}

\item The PDF transition density from lognormal to powerlaw is a mathematically motivated critical density and can be physically related to the density where the jeans length is comparable to the sonic length, i.e. the post-shock critical density for collapse. 

\item The transition density depends only on the properties of the PDF, such as the width and slope of the powerlaw, and therefore the calculation for the dense self-gravitating gas fraction and star formation efficiency is readily comparable with observables.

\item We test the analytic predictions for the transition density and self-gravitating gas fraction against {\sc AREPO} moving mesh gravoturbulent simulations and find good agreement.  

\item When the PDF transition density from powerlaw to lognormal forms is taken as the critical density for collapse, the instantaneous star formation efficiency can be calculated from the dense self-gravitating gas fraction represented as the fraction of gas in the powerlaw portion of the PDF.

\item Our results suggest that the self-gravitating gas fraction and the overall instantaneous SFE for a self-gravitating cloud with  $\alpha_{\rm vir}\approx 1$ should \textit{be slightly anti-correlated with sonic Mach number}. The depletion time increases with increasing sonic Mach number.

\item PAWS observations show a SFE which is slight anti-correlated with velocity dispersion (a proxy for the sonic Mach number), in agreement with the theory here.

\end{itemize}

\acknowledgments
B.B is thankful for valuable discussions with Hope Chen, Christoph Federrath, Alyssa Goodman, Mark Krumholz, Charlie Lada, Alex Lazarian, Adam Leroy, Christopher Mckee, Phil Myers, Anna Rosen,  Mohammad Safarzadeh, Zachary Slepian, Matthew Smith, and Catherine Zucker.   B.B. is also grateful for discussions with the SMAUG collaboration.
 B.B. acknowledges generous support from the Simons Foundation Center for Computational Astrophysics.  P.M. acknowledges support from NASA through Einstein Postdoctoral Fellowship grant number PF7-180164 awarded by the \textit{Chandra} X-ray Center, which is operated by the Smithsonian Astrophysical Observatory for NASA under contract NAS8-03060.  Both authors are grateful to the anonymous referee, whose comments and insights improved the paper dramatically. 
The computations in this paper were run on the Odyssey cluster supported by the FAS Division of Science, Research Computing Group at Harvard University.

\end{document}